\documentclass[preprint,superscriptaddress,amsmath,amssymb,aps,prb,floatfix]{revtex4-1}

\usepackage{graphicx}
\usepackage{enumitem}
\usepackage{gensymb}
\usepackage{amsmath}
\usepackage{amssymb}
\usepackage{dcolumn}
\usepackage{color}
\begin{document}

\title{A full configuration interaction quantum Monte Carlo study of ScO, TiO and VO molecules}

\author{Tonghuan Jiang}
\affiliation{School of Physics, Peking University, Beijing 100871, People’s Republic of China}

\author{Yilin Chen}
\affiliation{International Center for Quantum Materials, School of Physics, Peking University, Beijing 100871, People’s Republic of China}
\author{Nikolay Bogdanov}
\affiliation{Max Planck Institute for Solid State Research, Heisenbergstrasse 1, 70569 Stuttgart, Germany}
\author{Enge Wang}
\affiliation{International Center for Quantum Materials, School of Physics, Peking University, Beijing 100871, People’s Republic of China}

\affiliation{
 Collaborative Innovation Center of Quantum Matter, Beijing 100871, People’s Republic of China
}
\affiliation{Songshan Lake Materials Lab, Institute of Physics, Chinese Academy of Sciences, Guangdong, China.}

\author{Ali Alavi}
\affiliation{Max Planck Institute for Solid State Research, Heisenbergstrasse 1, 70569 Stuttgart, Germany}
\affiliation{University of Cambridge, Lensfield Road, Cambridge CB2 1EW, United Kingdom}

\author{Ji Chen}
\email{ji.chen@pku.edu.cn}
\affiliation{School of Physics, Peking University, Beijing 100871, People’s Republic of China}
\affiliation{
 Collaborative Innovation Center of Quantum Matter, Beijing 100871, People’s Republic of China
}
\affiliation{Max Planck Institute for Solid State Research, Heisenbergstrasse 1, 70569 Stuttgart, Germany}

\date{\today}

\begin{abstract}
Accurate \textit{ab initio} calculations of 3d transition metal monoxide molecules have attracted extensive attention because of its relevance in physical and chemical science, as well as theoretical challenges in treating strong electron correlation. 
Meanwhile, recent years have witnessed the rapid development of full configuration interaction quantum Monte Carlo (FCIQMC) method to tackle electron correlation.
In this study, we carry out FCIQMC simulations to ScO, TiO and VO molecules and obtain accurate descriptions of 13 low-lying electronic states (ScO $^2\Sigma^+$, $^2\Delta$, $^2\Pi$; TiO $^3\Delta$, $^1\Delta$, $^1\Sigma^+$, $^3\Pi$, $^3\Phi$; VO $^4\Sigma^-$, $^4\Phi$, $^4\Pi$, $^2\Gamma$, $^2\Delta$),
including states that have significant multi-configurational character.
The FCIQMC results are used to assess the performance of several other wave function theory and density functional theory methods.
Our study highlights the challenging nature of electronic structure of transition metal oxides and demonstrates FCIQMC as a promising technique going forward to treat more complex transition metal oxide molecules and materials.

\end{abstract}

\maketitle

\section{Introduction}
The 3d transition metal monoxide (TMO) molecules
have been the focus of electronic structure measurements and calculations,
which hold importance in theoretical understanding of chemical bonds in transition metal oxides\cite{Merer1989}. 
Additionally, these molecules play key roles in 
astrophysics \cite{Sedaghati2017, White1978, Merer1989}, catalytic reactions \cite{SHELDON1981152, Periana1998, Joergensen1989} and 
gas-phase chemistry \cite{Eller1991,Xu1998}. 
For example, 3d transition metal monoxides are symbolic molecules in M-type stars, due to the great stability of nuclei around $^{56}\text{Fe}$, the high cosmic abundance of oxygen, and large bonding energies
\cite{Sedaghati2017,McKemmish_TiO_2019}.
The observation and understanding of absorption spectra of these molecules are intimately related to their electronic structure.
As another example, 
in 
gas-phase chemistry, 
diatomic molecules are intermediate products in metal oxidation, 
and understanding of reaction mechanism requires knowledge of molecular bonding\cite{Xu1998}.
Therefore, it is of great importance to accurately describe the electronic structure of 3d transition metal monoxide molecules. 
In addition, a successful treatment of monoxide will provide a solid base for tackling more ambitious electron correlation problems in transition metal oxide materials, which remains a daunting challenge for  \textit{ab initio} calculations \cite{li_electronic_2019,williams_direct_2020,Chen_TiO2v_2020,katukuri_electronic_2020}.

Many calculations on these molecules have been reported using different electronic structure methods.
For example, Bridgeman and Rothery studied the bond lengths, bond energies and force constants of TMOs
using non-local density functional theory.\cite{Bridgeman2000}
Dai et al. calculated electron affinities and detachment energies of TMOs 
by means of time-dependent density functional theory. \cite{Dai2003}
Wagner and Mitas used fixed-node diffusion Monte Carlo(DMC) and reptation Monte Carlo to obtain the bond energies, bond lengths and dipole moments of the ground state of TMOs (ScO - MnO)
\cite{Wagner_dmc_2007}.
Miliordo et al. used internally contracted MRCI (icMRCI) method to calculate the potential curves and dipole moments of multiple low-lying electronic states of 
neutral and charged molecules
from ScO to NiO.\cite{Miliordos_VO_2007, Miliordos_ScOTiOCrOMnO_2010, Sakellaris_FeO_2011, Sakellaris_CoO_2012, Sakellaris_NiO_2013}
McKemmish et al. combined the results from MRCI calculations and  experimental data,
and produced potential curves of 13 low-energy states of TiO and VO molecules. \cite{McKemmish_TiO_2019, McKemmish_VO_2016}
%
Moltved and Kepp applied coupled cluster theory with singles, doubles and perturbative triples (CCSD(T)) to neutral molecules and monovalent cations of TMOs. \cite{Moltved_CC_2019}
Williams et al. recently compared the performance of a wide range of electronic structure methods on the ground states of transition metal atoms and molecules.\cite{williams_direct_2020}
%
%

Overall, these theoretical studies have revealed the great challenge to solve the electronic structure of TMOs, not to mention strongly correlated electronic states with varying spin multiplicity. \cite{Merer1989,Harrison2000}
Even for early 3d TMOs such as ScO, TiO and VO, theoretical descriptions remain unsatisfactory to some extent. \cite{Miliordos_VO_2007, McKemmish_VO_2016, Hopkins_2009, williams_direct_2020}
Regular mean-field methods, such as density functional theory (DFT), are generally considered not accurate enough to deal with strongly correlated excited states of TMO. 
Currently icMRCI is the most widely used method to deal with TMO\cite{Tennyson_openshell_2016}, but even here there are still some cases with large errors. 
For example, in Ref \onlinecite{Miliordos_ScOTiOCrOMnO_2010}, the icMRCI/aug-cc-pVQZ calculation overestimated excitation energy of TiO $^1\Sigma^+$ by $\sim1900\text{cm}^{-1}$. 
Although state-of-the-art calculations (such as AFQMC, SHCI, SEET, DMRG and i-FCIQMC) have been reported on TMOs' ground states \cite{williams_direct_2020}, excited states of TMOs have not been reported with these methods. 
Full configuration interaction quantum Monte Carlo (FCIQMC) provides a new approach to the accurate theoretical description of electronic structure. 
\cite{Booth_FCIQMC_2009}
FCIQMC solves the full configuration interaction problem with a stochastic sampling method.
The many body wave function is represented with evolution of `walkers' in the determinant Hilbert space. 
Despite its relatively high computational cost, FCIQMC with an initiator approximation, namely i-FCIQMC \cite{Cleland_iFCIQMC_2010}, has been successfully applied to a variety of systems such as first row diatomics \cite{Cleland_firstrow_2012}, a number of molecules from the G1 standard set\cite{Kersten_ExC&G1_2016}, first row transition metal atoms \cite{Thomas_metal_2015}, ground states of TMOs \cite{williams_direct_2020}, and simple solids\cite{Booth_solid_2013}. 
Recently, the newly developed {\em adaptive shift} method (AS-FCIQMC) has further improved the accuracy and efficiency of FCIQMC\cite{Ghanem_adapshift_2019,ghanem_adaptive_2020}.
Therefore, it is desirable to investigate challenging TMO excited states with AS-FCIQMC.
In this article, we report AS-FCIQMC calculations on ScO, TiO and VO molecules. 
13 low-lying electronic states (ScO $^2\Sigma^+$, $^2\Delta$, $^2\Pi$; TiO $^3\Delta$, $^1\Delta$, $^1\Sigma^+$, $^3\Pi$, $^3\Phi$; VO $^4\Sigma^-$, $^4\Phi$, $^2\Gamma$, $^4\Pi$, $^2\Delta$) are computed in total.
We reveal the strongly multi-configurational nature of the excited states computed, namely their wave functions are dominated by not single but multiple Slater determinants.
%
The dissociation energies, bond lengths, vibrational frequencies, and excitation energies obtained from our calculations are compared with experimental results available.
In addition, FCIQMC results can be used to further assess other commonly used electronic structure methods.

\section{Methods and computational details}

FCIQMC calculation is performed with the NECI code. \cite{Booth_NECI_2014,Kai_NECI_2020}
Adaptive shift approach (AS-FCIQMC)\cite{Ghanem_adapshift_2019, ghanem_adaptive_2020} is adopted in all FCIQMC calculations, with initiator threshold $n_a=3$. 
Different adaptive shift offsets have been introduced to validate the convergence of AS-FCIQMC.\cite{ghanem_adaptive_2020}
The time-step is iteratively updated to keep the calculation stable. 
Semi-stochastic method\cite{Blunt_semistoch_2015} is used, and the size of deterministic space is set to 100. 
12 core electrons are frozen ($1s^22s^22p^6$ of M and $1s^2$ of O) to reduce the dimension of FCI space involved. 
Trial wave function is used to obtain the projection energy estimation, and the size of trial space $\mathcal{T}$ is set to 10, unless otherwise specified. 
Except for TiO $^1\Sigma^+$ and all excited states of ScO and VO which are evaluated with $L_z$ symmetry\cite{Booth_Lz_2011}, all low-lying states are calculated with $C_{2v}$ symmetry. 
For the two low-spin states with even open shells, $\text{TiO} ^1\Delta$ and $^1\Sigma^+$, half-projected Hartree-Fock functions are used\cite{Cox_HPHF_1976}. 
The Dunning type correlation-consistent basis sets at triple zeta level (cc-pVTZ) are used for oxygen and metal atoms Sc, Ti and V\cite{BasisSetExchange_2019}.
The basis set contains basis functions of 7s6p4d2f1g (68 functions) for metal atoms involved, and 4s3p2d1f (30 functions) for oxygen atom. 
Molecular orbitals are obtained via non-relativistic restricted open-shell Hartree-Fock (ROHF) calculations with PySCF 1.6.4 package\cite{PySCF}.

The total energies of the ground state and excited states of ScO, TiO and VO molecules are evaluated at ground state equilibrium geometries.
The vertical excitation energy is calculated, and then compared with literature results.
The literature vertical excitations are obtained by adding up two parts: $\Delta E_\text{v}=T_\text{e}+\Delta E$.
$T_\text{e}$ is the adiabatic energy separation from experiment.
$\Delta E$ is derived from harmonic ($\omega_e$) and anharmonic ($\omega_e x_e$) frequencies from literature by assuming the 3rd-order anharmonicity:
\begin{equation}
    \Delta E=\frac{1}{2}m\omega_e^2(r-r_e)^2-\frac{1}{6}m\omega_e^2\eta_3(r-r_e)^3 \label{U^(3)}
\end{equation}
where m is the reduced mass of the molecule. 
The anharmonic frequency $\omega_e x_e$ is defined with the vibration spectrum of the anharmonic oscillator: 
\begin{equation}
    E_n=\hbar\omega_e(n+\frac{1}{2})-\hbar\omega_e x_e(n+\frac{1}{2})^2, n=0,1,2,... \label{E_nu}
\end{equation}
From the 2nd-order perturbation calculation, the anharmonic frequency can be associated with $\eta_3$ as follows:
\begin{equation}
    \eta_3=\sqrt{\frac{12m\omega_e x_e}{5\hbar}} \label{eta_3}
\end{equation}
The total energies of Sc, Ti, V and O atoms are also evaluated via FCIQMC, in which 10 core electrons in metal atoms and 2 in oxygen  are frozen.
The dissociation energies are evaluated via the equaiton $D_\text{e} = E(\text{M})+E(\text{O})-E(\text{MO})$, M=(Sc, Ti, V). 
For atoms, $D_{2h}$ symmetry is adopted, and the irreducible representation (irrep) of atomic states (Sc, $^2D$; Ti, $^3F$; V, $^4F$; O, $^3P$) are chosen according to the adiabatic connection rules discussed in Refs. \onlinecite{Miliordos_ScOTiOCrOMnO_2010} and \onlinecite{Miliordos_VO_2007}. 
The bonding curves of the ground states of ScO, TiO and VO are plotted 

and fitted to the Morse potential (Eq. \ref{Morse_potential}) \cite{Morse_potential_1929}.
\begin{equation}
    V(r)=D_\text{e}(1-e^{-a(r-r_\text{e})})^2 \label{Morse_potential}
\end{equation}
where $D_\text{e}$, the dissociation energy, is obtained by $E(\text{M})+E(\text{O})-E(\text{MO})$. 
%

Vertical excitation energies of low-lying states of ScO, TiO and VO are obtained with other computational methods implemented in PySCF 1.6.4, including wave function theory (WFT) and DFT.
The WFT methods include Hartree-Fock, MP2 (second order M{\o}ller-Plesset perturbation theory), CCSD (coupled cluster theory with singles and doubles) and CCSD(T) (coupled cluster theory with singles, doubles and perturbative triples correction).
B3LYP\cite{Stephens_B3LYP_1994}, PBE0\cite{Adamo_PBE0_1999}, PBE\cite{Perdew_PBE_1996}, PW91\cite{Perdew_PW91_1992} and LDA exchange correlation functionals are used in DFT calculations.
In all calculations other than FCIQMC, no core electrons are frozen.

\section{Results}

\subsection{Excitation energies}

\subsubsection{ScO}
\begin{figure}
    \centering
    \includegraphics[width=8.0cm]{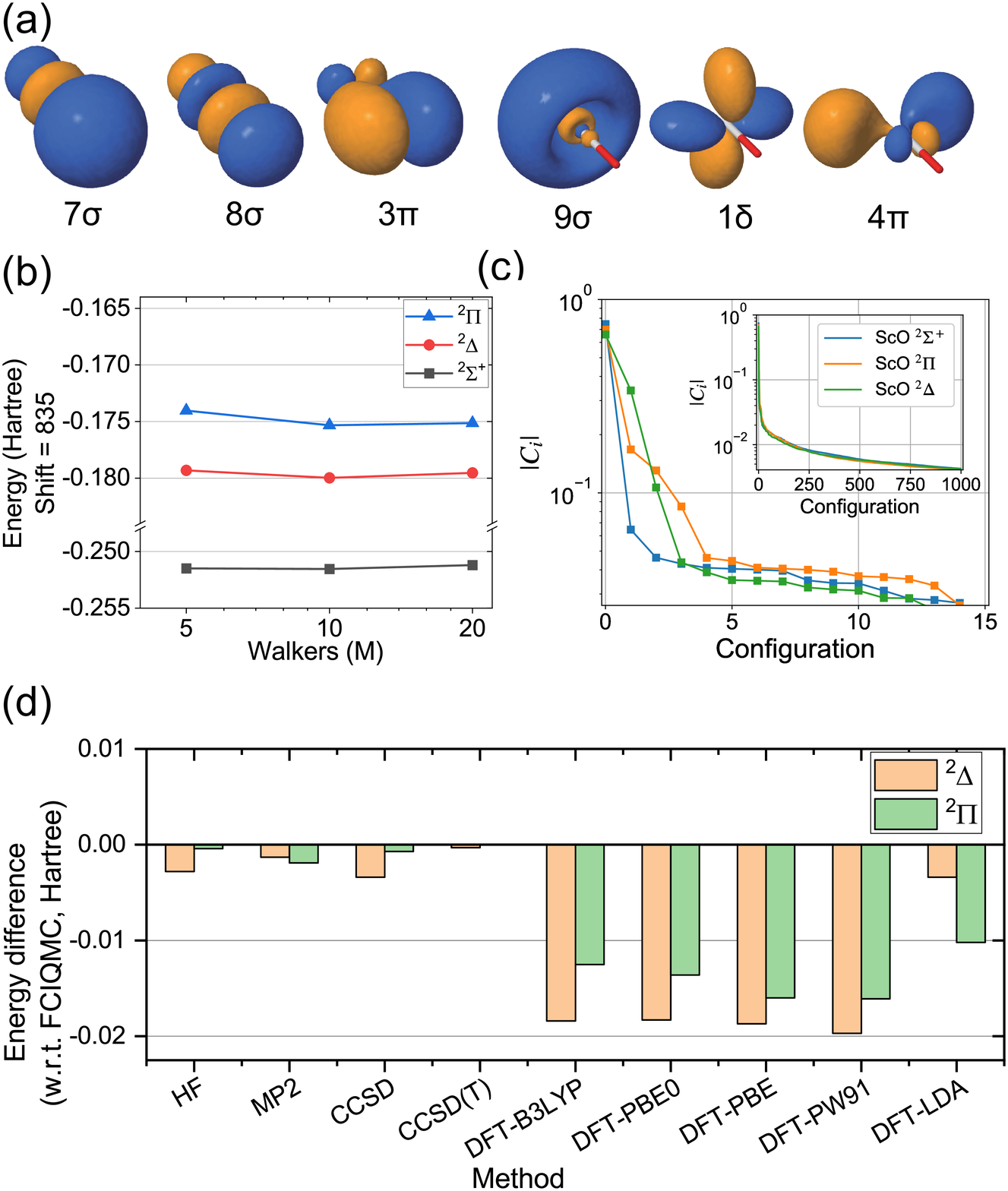}
    \caption{The results of ScO. 
    (a) Six frontier molecular orbitals of ScO, including $7\sigma$, $8\sigma$, $3\pi$, $9\sigma$, $1\delta$ and $4\pi$. 
    (b) Projection energies of the low-lying states, $^2\Sigma^+$, $^2\Delta$ and $^2\Pi$, as a function of the total walker number. All curves in this figure are shifted upwards by $+835\text{E}_\text{h}$. 
    (c) Amplitude of the 15 (in main panel) and 1000 (in inset) most populated configurations in the wave function from FCIQMC calculations on low-lying states at ground state equilibrium geometry. All curves are obtained with 20M walkers. 
    (d) The difference in vertical excitation energy between FCIQMC and other computational methods. The positive difference indicates that the method obtains a vertical excitation energy higher than that of FCIQMC. 
    }
    \label{Fig_ScO}
\end{figure}

A ScO molecule contains 29 electrons. 
18 core electrons of Sc form a closed shell of Sc[Ar], and 2 core electrons of O form a closed shell of O[He].
The remaining valence electrons occupy the frontier molecular orbitals shown in Fig. \ref{Fig_ScO}(a). 
The $7\sigma$ and $8\sigma$ orbitals are two bonding orbitals formed by Sc $3d_{z^2}$, O $2s$ and O $2p_z$. 
The two-fold degenerate orbitals, $3\pi_x$ and $3\pi_y$, are bonding orbitals from (Sc $3d_{xz}$, O $2p_x$) and (Sc $3d_{yz}$, O $2p_y$), respectively. 
These four orbitals are doubly occupied by 8 electrons. 
$9\sigma$, $1\delta$ and $4\pi$, however, come from Sc $4s$, Sc $3d_{x^2-y^2}$ and Sc $3d_{xz}$ respectively. 
These three orbitals are singly occupied in the three lowest-lying states of ScO, $^2\Sigma^+$, $^2\Delta$ and $^2\Pi$ respectively. 
The ground state is $^2\Sigma^+$.
The FCIQMC projection energies of ScO's low-lying states as a function of the number of walkers are plotted in Fig. \ref{Fig_ScO}(b).
For all the states studied, 
energy converges within $0.5\text{mE}_\text{h}$ as the walker number grows from 10M to 20M.

The relatively fast convergence of energy with respect to the walker number demonstrates the efficiency of AS-FCIQMC.
In Fig. \ref{Fig_ScO}(c) we show 
the amplitudes of 15 highest populated configurations in the determinant space.
All the dominant configurations of the three low-lying states have an amplitude  about 0.7. 
The multi-configurational character becomes increasingly significant from $^2\Sigma^+$ to  $^2\Pi$ to $^2\Delta$, with $^2\Delta$ having an amplitude of 0.35 on the second populated determinant.
%

Table \ref{ScO_vertical} presents the total energy and vertical excitation energy of low-lying states of ScO at the equilibrium geometry $r_\text{e}=1.6682\text{\AA}$, along with experimental results in literature\cite{Chalek_ScO2D_1976, Chalek_ScO2P_1977}.
Our results of $^2\Delta$ and $^2\Pi$ well reproduce the experimental results with an error of only $1.5\text{mE}_\text{h}$ and $0.8\text{mE}_\text{h}$ respectively.
This highlights the accuracy of FCIQMC in these systems.

\begin{table}[]
    \centering
    \begin{tabular}{|p{1cm}<{\centering}|p{3cm}<{\centering}|p{3cm}<{\centering}|p{2.5cm}<{\centering}|p{3cm}<{\centering}|}
    \hline
        State & Configuration & Total energy ($\text{E}_\text{h}$) & $\Delta E_\text{v}$($\text{cm}^{-1}$) & Exp. ($\text{cm}^{-1}$)\\
    \hline
        $^2\Sigma^+$ & $9\sigma^1$ & -835.25120(14) & 0 & 0\\
        $^2\Delta$ & $1\delta^1$ & -835.17952(13) & 15730(43) & 15394.7 ($^2\Delta_{3/2}$) 15500.1 ($^2\Delta_{5/2}$)$^a$ \\
        $^2\Pi$ & $4\pi^1$ & -835.17513(18) & 16694(50) & 16523.5 ($^2\Pi_{1/2}$) 16655.6 ($^2\Pi_{3/2}$)$^b$ \\
    \hline
    \end{tabular}
    \caption{Total energy and vertical excitation energy of three lowest-lying states of ScO at its ground state equilibrium bond length ($r_\text{e}=1.668\text{\AA}$)\cite{Harrison2000}. 
    The literature data are obtained via equation $\Delta E_\text{v}=T_\text{e}+\Delta E$, 
    where $\Delta E$ is calculated with Eq \ref{U^(3)} and \ref{eta_3}. 
    $T_\text{e}$, $\omega_e$, $\omega_e x_e$ are experiment results as follows:
    $^a$ $T_\text{e}$, Ref \onlinecite{Chalek_ScO2D_1976}, chemiluminescence spectroscopy; $\omega_e$ and $\omega_e x_e$, Ref \onlinecite{Rice_ScO2D_1989}, fluorescence excitation spectroscopy. 
    $^b$ $T_\text{e}$, $\omega_e$ and $\omega_e x_e$ all from Ref \onlinecite{Chalek_ScO2P_1977}, single collision chemiluminescence study. }
    \label{ScO_vertical}
\end{table}

With FCIQMC results as benchmark, now we can discuss the performance of other WFT and DFT methods without addressing effects beyond electronic structure, as is shown in Fig. \ref{Fig_ScO}(d).
CCSD(T) performs quite well for the three ScO states with a minor error of $0.2\text{mE}_\text{h}$ compared with FCIQMC.
CCSD(T) is known as the golden standard for single-reference electronic structure calculations, and our calculations show that it can also be accurate for systems with some significant multi-configurational character.
CCSD without perturbative triples, however, has a much larger error of $2.4\text{mE}_\text{h}$.
Surprisingly, Hartree-Fock calculations have resulted in almost the same accuracy as that of CCSD and MP2 for ScO.
DFT methods generally yield huge errors \textgreater $10\text{mE}_\text{h}$, and there is no obvious difference from one functional to another.
%
It is worth noting that although LDA appears to give the smallest errors among all functionals considered, the energy ordering of the $^2\Delta$ and $^2\Pi$ states is reversed.

\subsubsection{TiO}

\begin{figure}
    \centering
    \includegraphics[width=8.0cm]{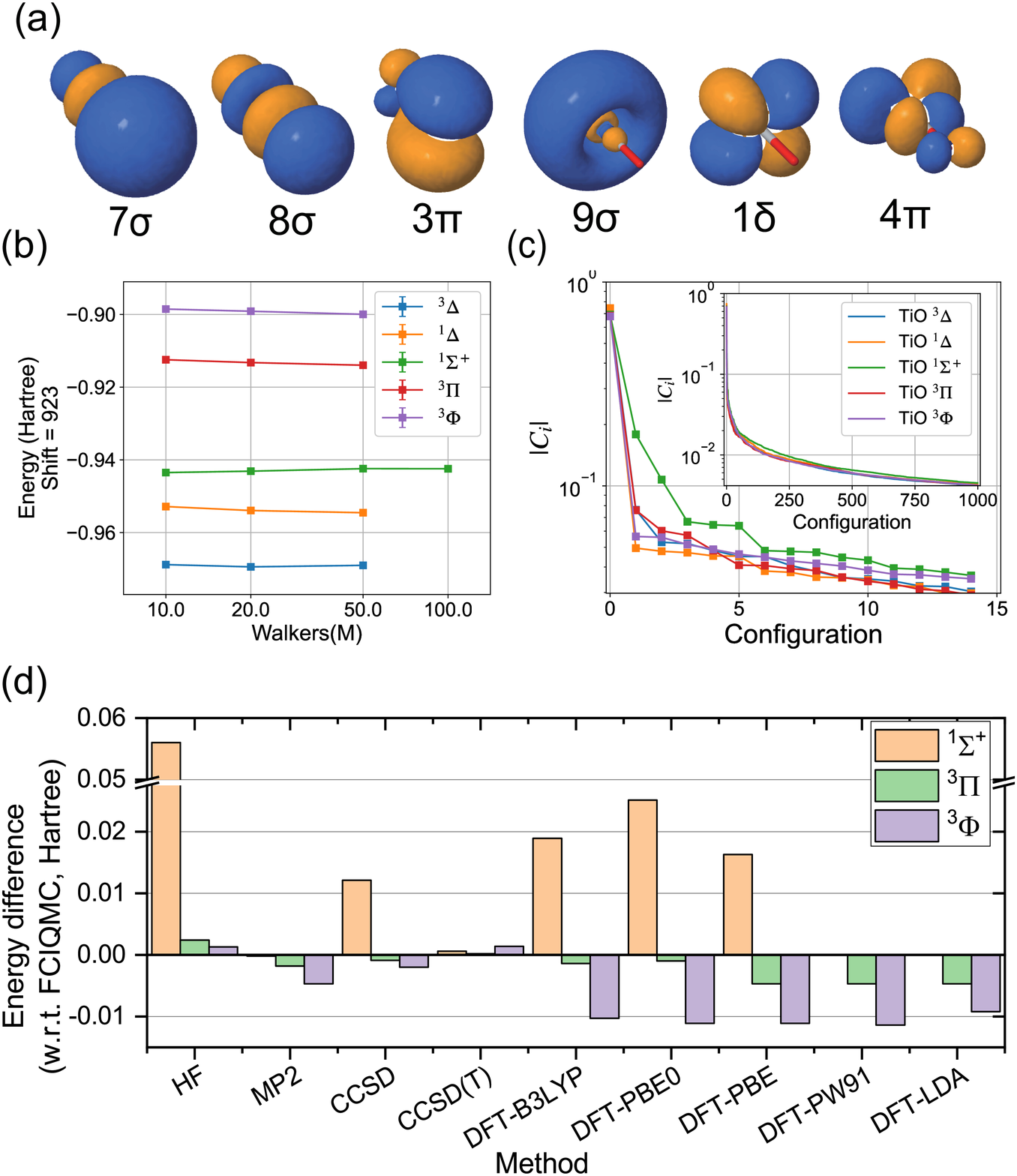}
    \caption{The results of TiO. 
    (a) Six frontier molecular orbitals of TiO, including $7\sigma$, $8\sigma$, $3\pi$, $9\sigma$, $1\delta$ and $4\pi$. 
    (b) Projection energies of the low-lying states, $^3\Delta$,  $^1\Delta$, $^1\Sigma^+$, $^3\Pi$ and $^3\Phi$, as a function of the total walker number. $^1\Sigma^+$ curve is obtained with adaptive-shift offset = $E_\text{ref}$. All curves in this figure is shifted upwards by $+923\text{E}_\text{h}$.
    (c) Amplitude of the 15 (in main panel) and 1000 (in inset) most populated configurations in the wave function from FCIQMC calculations on low-lying states at ground state equilibrium geometry. 
    (d) The difference in the vertical excitation energy between FCIQMC and other \textit{ab initio} methods. A positive number indicates that the method obtains a higher vertical excitation energy than FCIQMC. The difference of CCSD(T) is too small to be seen in the figure. DFT-PW91 and DFT-LDA fail to converge for TiO $^1\Sigma^+$.
    }
    \label{Fig_TiO}
\end{figure}

A TiO molecule contains 30 electrons.
The source of orbitals in TiO are similar to ScO. 
The only difference is that it has 2 electrons occupying higher molecular orbitals, namely $9\sigma$, $1\delta$, and $4\pi$, as shown in Fig. \ref{Fig_TiO}(a) along with other valence orbitals. 
%
The FCIQMC projection energies of TiO's low-lying states (three triplet states and two singlet states) are plotted in Fig. \ref{Fig_TiO}(b) as a function of the number of walkers.
In general, TiO states are more complex than the ScO states, and we need at least 50M walkers to reach a converge of  $1.2\text{mE}_\text{h}$ for their energies.
In addition, we find that the singlet states ($^1\Delta$ and $^1\Sigma^+$) converge relatively slower than the triplet states ($^3\Delta$, $^3\Pi$, $^3\Phi$), requiring a larger number of walkers.

\begin{table}[]
    \centering
    \begin{tabular}{|p{1cm}<{\centering}|p{3cm}<{\centering}|p{3cm}<{\centering}|p{2.5cm}<{\centering}|p{3cm}<{\centering}|p{1.5cm}<{\centering}|}
    \hline
        State & Configuration & Total energy ($\text{E}_\text{h}$) & $\Delta E_\text{v}$ ($\text{cm}^{-1}$)  & Ref ($\text{cm}^{-1}$) \\
    \hline
        $^3\Delta$ & $9\sigma^11\delta^1$ & -923.96873(15) & 0 & 0\\
        $^1\Delta$ & $9\sigma^11\delta^1$ & -923.95342(18) & 3361(52)  & 3446 \\
        $^1\Sigma^+$ & $9\sigma^2$ & -923.94244(17) & 5771(50) &  5741 \\
        $^3\Pi$ & $9\sigma^14\pi^1$ & -923.91333(24) & 12160(62) &  12052 \\ 
        $^3\Phi$ & $1\delta^14\pi^1$ & -923.89920(15) & 15261(47) &  14363 \\
    \hline
    \end{tabular}
    \caption{Total energy and vertical excitation energy of five low-lying states of TiO at its ground state equilibrium bond length ($r_\text{e}=1.620\text{\AA}$)\cite{Harrison2000}. 
    The literature data are obtained via equation $\Delta E_\text{v}=T_\text{e}+\Delta E$, 
    where $\Delta E$ is calculated with Eq \ref{U^(3)} and \ref{eta_3}.
    Source of $T_\text{e}$, $\omega_e$ and $\omega_e x_e$:
    $^1\Delta$, $T_\text{e}$, Ref \onlinecite{Amiot_TiO1D_1996}, laser induced fluorescence (LIF) spectroscopy; $\omega_e$ and $\omega_e x_e$, Ref \onlinecite{Brandes_TiO1D_1985}, Fourier transform(FT) rovibrational analysis.
    $^1\Sigma^+$, $T_\text{e}$, Ref \onlinecite{Kaledin_TiO1S+_1995}, LIF spectroscopy; $\omega_e$ and $\omega_e x_e$, Ref \onlinecite{Galehouse_TiO1S+_1980}, Fourier transform near infrared (near FTIR) spectroscopy.
    $^3\Pi$, $T_\text{e}$, Ref \onlinecite{Simard_TiO3P_1991}, LIF spectroscopy; $\omega_e$, Ref \onlinecite{Kobayashi_TiO_3P_2002}, frequency modulated laser absorption spectroscopy; $\omega_e x_e$, Ref \onlinecite{Miliordos_VO_2007}, MRCI+Q calculations. 
    $^3\Phi$, $T_\text{e}$, Ref \onlinecite{Barnes_TiO3F_1996}, double resonance spectroscopy; $\omega_e$ and $\omega_e x_e$, Ref \onlinecite{Ram_1999} and \onlinecite{Amiot_2002}, FT spectroscopy and LIF spectroscopy.
    }
    \label{TiO_vertical}
\end{table}

The FCIQMC total energies and vertical excitation energies are shown in Table \ref{TiO_vertical}, in comparison with experimental results.
%
At the equilibrium geometry $r_\text{e}=1.620\text{\AA}$, the ground state $^3\Delta$ has a total energy of $-923.96873(15)\text{E}_\text{h}$.
%
%
The vertical excitation energy of $^1\Delta$, $^1\Sigma^+$ and $^3\Pi$ all show good agreement with literature, with differences all below $0.5\text{mE}_\text{h}$.
The only exception is the $^3\Phi$, which has a much higher excitation energy than experiment ($\sim900\text{cm}^{-1}$; $4.1\text{mE}_\text{h}$).
There are multiple possible sources of this error, such as basis set incompleteness error, relativistic effect, frozen core approximation, etc. 
These effects are beyond the scope of this study, but it is worth to mention that such effects would affect the assessment of theoretical results compared with experiment. \cite{williams_direct_2020}

In Fig. \ref{Fig_TiO} (d) FCIQMC are compared with WFT and DFT methods. 
%
Overall, the performance of these methods are similar to what have been discussed for the ScO molecule.
Except for HF which suffers from a large deviation in the $^1\Sigma^+$ state, WFT generally has smaller errors than DFT.
%
%

\subsubsection{VO}

\begin{figure}
    \centering
    \includegraphics[width=8.0cm]{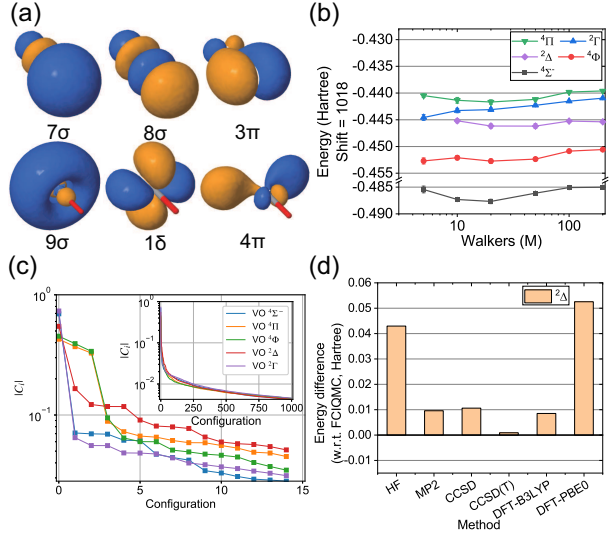}
    \caption{The results of VO. 
    (a) Six frontier molecular orbitals of VO, including $7\sigma$, $8\sigma$, $3\pi$, $9\sigma$, $1\delta$ and $4\pi$. 
    (b) Projection energies of the low-lying states, $^4\Sigma^-$,  $^4\Phi$, $^2\Gamma$, $^4\Pi$ and $^2\Delta$, as a function of the total walker number. The curves in this figure are shifted upwards by $1018\text{E}_\text{h}$. $^4\Sigma^-$ curve is obtained with adaptive-shift offset = $E_\text{ref}$.
    (c) Amplitude of the 15 (in main panel) and 1000 (in inset) most populated configurations in the wave function from FCIQMC calculations on low-lying states at ground state equilibrium geometry. 
    (d) The difference in vertical excitation energy between FCIQMC and other computational methods. A positive number indicates a higher vertical excitation energy than FCIQMC. PBE, PW91 and LDA fail to converge for VO $^2\Delta$.
    }
    \label{Fig_VO}
\end{figure}

\begin{table}[]
    \centering
    \begin{tabular}{|p{1cm}<{\centering}|p{2.5cm}<{\centering}|p{3cm}<{\centering}|p{2.2cm}<{\centering}|p{1.5cm}<{\centering}|p{1.5cm}<{\centering}|p{1.5cm}<{\centering}|p{1.5cm}<{\centering}|}
    \hline
        State & Configuration & Total energy ($\text{E}_\text{h}$) & $\Delta E_\text{v}$ ($\text{cm}^{-1}$) & Ref $^a$ ($\text{cm}^{-1}$) & Ref $^b$ ($\text{cm}^{-1}$) \\
    \hline
        $^4\Sigma^-$ & $9\sigma^11\delta^2$ & -1018.48501(14) & 0 & 0 & 0 \\
        $^4\Phi$ & $9\sigma^11\delta^14\pi^1$ & -1018.45057(19) & 7559(52) & 7515 & 7511  \\
        $^2\Gamma$ & $9\sigma^11\delta^2$ & -1018.44095(13) & 9670(42) & \textemdash & \textemdash \\
        $^4\Pi$ & $9\sigma^11\delta^14\pi^1$ & -1018.43959(18) & 9967(50) & 9851 & 9832  \\ 
        $^2\Delta$ & $9\sigma^21\delta^1$ & -1018.44538(23) & 8698(60) & 9924 & 9378(90)  \\
    \hline
    \end{tabular}
    \caption{Total energy and vertical excitation energy of five low-lying states of VO at its ground state equilibrium bond length ($r_\text{e}=1.589\text{\AA}$)\cite{Harrison2000}. 
    The literature data are obtained via equation $\Delta E_\text{v}=T_\text{e}+\Delta E$, 
    where $\Delta E$ is calculated with Eq \ref{U^(3)} and \ref{eta_3}.
    Source of $T_\text{e}$:
    $^a$ Ref \onlinecite{Hubner_VO_2015}, excitation energy from Ne-matrix isolation. 
    $^b$ Experimental excitation energy from other literature: 
    $^4\Phi$ (Ref \onlinecite{Merer_VO_1987}) and $^4\Pi$ (Ref \onlinecite{Cheung_VO4P_1982}) are from high-resolution Fourier transform spectroscopy; 
    $^2\Delta$, Ref \onlinecite{Ram_VO2D_2002, Ram_VO2D_2005}, Fourier transform emission spectroscopy.
    Source of $\omega_e$: Ref \onlinecite{Pedley_MO_1983}, high-resolution Fourier transform spectroscopy.
    Source of $\omega_e x_e$: Ref \onlinecite{Miliordos_VO_2007}, MRCI+Q calculations. 
    }
    \label{VO_vertical}
\end{table}

A VO molecule has 31 electrons, and the higher orbitals (Fig. \ref{Fig_VO}(a)) are occupied by 3 electrons. 
The 3 electrons give rise to multiple low-lying states, among which the ground state is $^4\Sigma^-$. 
The FCIQMC projection energies of five low-lying states of VO are plotted in Fig. \ref{Fig_VO} (b) as a function of walker number. 
All the states studied except $^4\Sigma^-$ have reached convergence within $1\text{mE}_\text{h}$ with a walker number up to 50M.
For $^4\Sigma^-$, the energy converges within $1.1\text{mE}_\text{h}$ when the walker number grows from 50M to 100M.

The amplitudes of 15 most populated determinants are shown in Fig. \ref{Fig_VO}(c). 
We find $^4\Sigma^-$ and $^2\Gamma$ are relatively single-configurational.
The amplitude of the most populated configuration is about 0.7, while that of the second most populated configuration is lower than 0.1.
$^4\Phi$ and $^4\Pi$ both have three highly populated Slater determinants with amplitude larger than 0.3.
%
$^2\Delta$ is also a multi-configurational state, with amplitude of 5 configurations larger than 0.1.

The vertical excitation energies are shown in Table \ref{VO_vertical} along with experimental results from literature\cite{Hubner_VO_2015, Merer_VO_1987}. 
The ground state $^4\Sigma^-$ has a total energy of $-1018.48501\text{E}_\text{h}$. 
The results of two quartets, $^4\Phi$ and $^4\Pi$ agree well with experimental results within $0.7\text{mE}_\text{h}$. 
Experimental result for the $^2\Gamma$ state is lacking, and our calculation shows it is about $9670\text{cm}^{-1}$.
%
For the $^2\Delta$ state, the theoretical and experimental results still have large discrepancy, which should be further investigated in the future.

We have computed the excitation energy of the $^2\Delta$ state and compared it with other WFT and DFT methods in Fig. \ref{Fig_VO}(d). 
Compared with ScO and TiO, CCSD(T) yields slightly larger error for the vertical excitation energy of VO $^2\Delta$ ($1.6\text{mE}_\text{h}$) due to its multi-configurational nature.
The deviation of CCSD from FCIQMC is one order of magnitude large than CCSD(T) 
($11\text{mE}_\text{h}$).
Among DFT methods, B3LYP provides more accurate excitation energy than PBE0, while PBE, PW-91 and LDA functionals fail to converge for the $^2\Delta$ state of VO. 

\subsection{Adaptive-shift offset}

\begin{figure}
    \centering
    \includegraphics[width = 8.0cm]{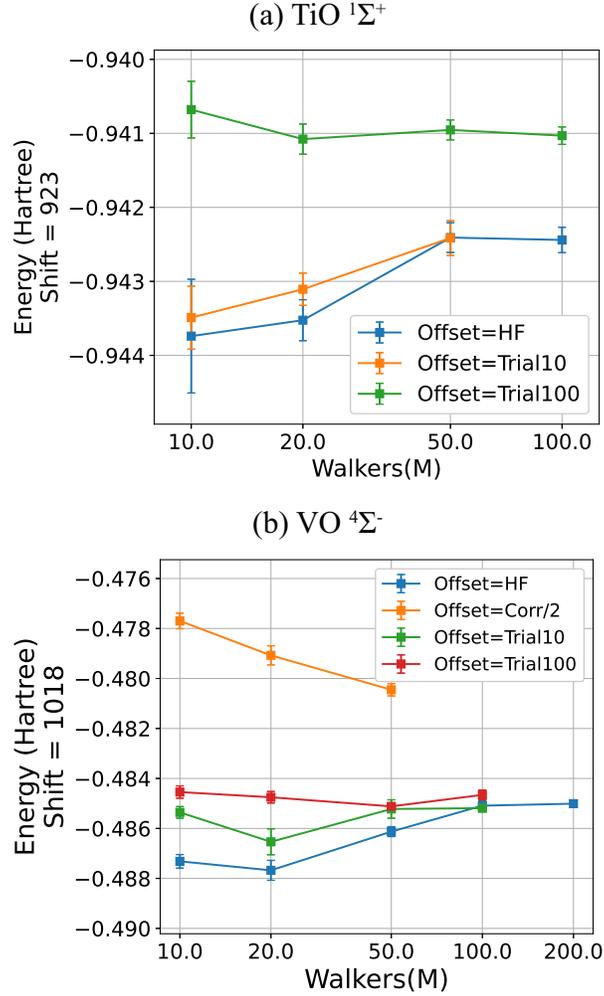}
    \caption{Projection energy of (a) TiO $^1\Sigma^+$ and (b) VO $^4\Sigma^-$ as a function of the total number of walkers, with different adaptive-shift offsets used. HF: $E_\text{ref}$; Corr/2: $(E_\text{ref}+E_0)/2$; ``Trial10": $E_\text{trial}$ with the size of trial space $\mathcal{T}=10$; ``Trial100": $E_\text{trial}$ with the size of trial space $\mathcal{T}=100$. 
    }
    \label{Fig_offset}
\end{figure}

The adaptive-shift offset (as-offset)\cite{ghanem_adaptive_2020} is an important parameter that affects the convergence of projection energy in AS-FCIQMC calculations. In this method, the shift $S_i$ applied to a non-initiator determinant is determined as follows:
\begin{equation*}
    S_i = \Delta + f_i\times(S - \Delta)
\end{equation*}
where $S$ is the global shift, $\Delta$ is the adaptive-shift offset, and $f_i$ is the scaling parameter for the non-initiator determinant, determined by the acceptance probability of a spawn from that determinant (See Eq 12 of Ref \onlinecite{ghanem_adaptive_2020}). The $\Delta$ parameter does not affect the converged energy, but influences the rate of convergence with respect to walker number. In the original adaptive shift the $\Delta$ parameter was set to the Hartree-Fock energy, whilst in Ref \onlinecite{ghanem_adaptive_2020} it was found that a lower value, typically by half of the correlation energy, was near optimal. Here we investigated several values of the $\Delta$, namely the HF energy , the energy of trial wavefunctions with 10 and 100 important determinants ("Trial10" and "Trial100" respectively), and HF + half of the estimated exact correlation energy ("corr/2").  
Fig. \ref{Fig_offset} shows the results of TiO $^1\Sigma^+$ and VO $^4\Sigma^-$ from different adaptive-shift offsets (see caption for details of offset).
The consistent converging pattern obtained from ``HF" and ``Trial10" further validate the convergence of the $^1\Sigma^+$ state (Fig. \ref{Fig_offset}(a)).
We note that although a good choice of as-offset leads to a faster convergence and a cross-check of the results, an unreasonable choice would lead to slower convergence.
For example, when we use the ``Trial100" offset, the convergence is not reached at 100M walkers for the $^1\Sigma^+$ state of TiO.

For the $^4\Sigma^-$ state of VO, the analysis of determinant population brings an interesting finding that the least multi-configurational wavefunction requires the largest walker number to converge.
Therefore, we further test the effect of as-offset on VO $^4\Sigma^-$  in Fig. \ref{Fig_VO}(b). 
In this case, very different convergence behavior is observed when different as-offset is applied.
Although the``HF",``Trial10" and ``Trial100" offsets all converge to the same result within 100M walkers, the ``Trial100" has a much faster convergence pattern.
%
%
We also tested the VO $^4\Sigma^-$ state the "Corr/2" offset, in which half of the correlation energy is taken as the offset.
However, it leads to a worse convergence behavior of AS-FCIQMC for the VO $^4\Sigma^-$ state.

\subsection{Bonding properties}

\begin{table}[]
    \centering
    \begin{tabular}{|c c|c|c|c|c|c|}
    \hline
         & & $D_\text{e}(\text{E}_\text{h})$ & $a  (\text{\AA}^{-1})$ & $r_\text{e} (\text{\AA})$ & $E_\text{eq} (\text{E}_\text{h})$ & $\omega (\text{cm}^{-1})$  \\
    \hline
        ScO & This work & 0.2490(1) & 1.654(21) & 1.6701(63) & -835.2453(20) & 925(12)\\
        & Exp. & 0.2543(3)$^a$ & \textemdash & 1.6656$^b$ & \textemdash & 975.7$^c$ \\
        TiO & This work & 0.2474(2) & 1.854(26) & 1.6120(37) & -923.9681(14) & 1024(15)\\
        & Exp. & 0.2524(26)$^d$ & \textemdash & 1.6203$^e$ & \textemdash & 1009$^e$ \\
        VO & This work & 0.2353(2) & 1.853(18) & 1.5913(34) & -1018.4844(11) & 991(10) \\
        & Exp. & 0.2382(32)$^f$ & \textemdash & 1.589$^g$ & \textemdash & 1002$^g$ \\
    \hline
    \end{tabular}
    \caption{Morse parameters and their derived quantities of ScO $^2\Sigma^+$, TiO $^3\Delta$ and VO $^4\Sigma^-$ from FCIQMC calculation and experiment, including the dissociation energy $D_\text{e}$, the equilibrium bond length $r_\text{e}$, the total energy at equilibrium geometry $E_\text{eq}$, and the harmonic frequency $\omega$. $^a$ Ref \onlinecite{Luc_ScO_2001, Jeung_ScO_2002}, cw laser-induced fluorescence spectroscopy(LIF); $^b$, Ref \onlinecite{Liu_ScO_1977}, LIF; $^c$, Ref \onlinecite{Chalek_ScO2P_1977}, single-collision chemiluminescence study; $^d$ $D_\text{e}$ is obtained indirectly by $D_\text{e}(\text{TiO})=D_0(\text{TiO}^+)+\text{IE}(\text{TiO})-\text{IE}(\text{Ti})$, where IE(TiO) (IE = ionization energy) is taken from two-color photoionization efficiency (PIE) spectroscopy and mass-analyzed threshold ionization (MATI)\cite{Loock_IE_1998}, and $D_0(\text{TiO}^+)$ is obtained from guided ion-beam mass spectroscopy\cite{Clemmer_D0_1991}; $^e$ Ref \onlinecite{Amiot_2002, Ram_1999}, Fourier transform (FT) spectroscopy and laser induced fluorescence spectroscopy (LIF); $^f$, Ref \onlinecite{Lagerqvist_1956, Lagerqvist_1957}, emission spectroscopy; $^g$, Ref \onlinecite{Balducci_1983}, high temperature-mass spectrometry.}
    \label{Morse_param}
\end{table}

\begin{figure}
    \centering
    \includegraphics[width = 8.0cm]{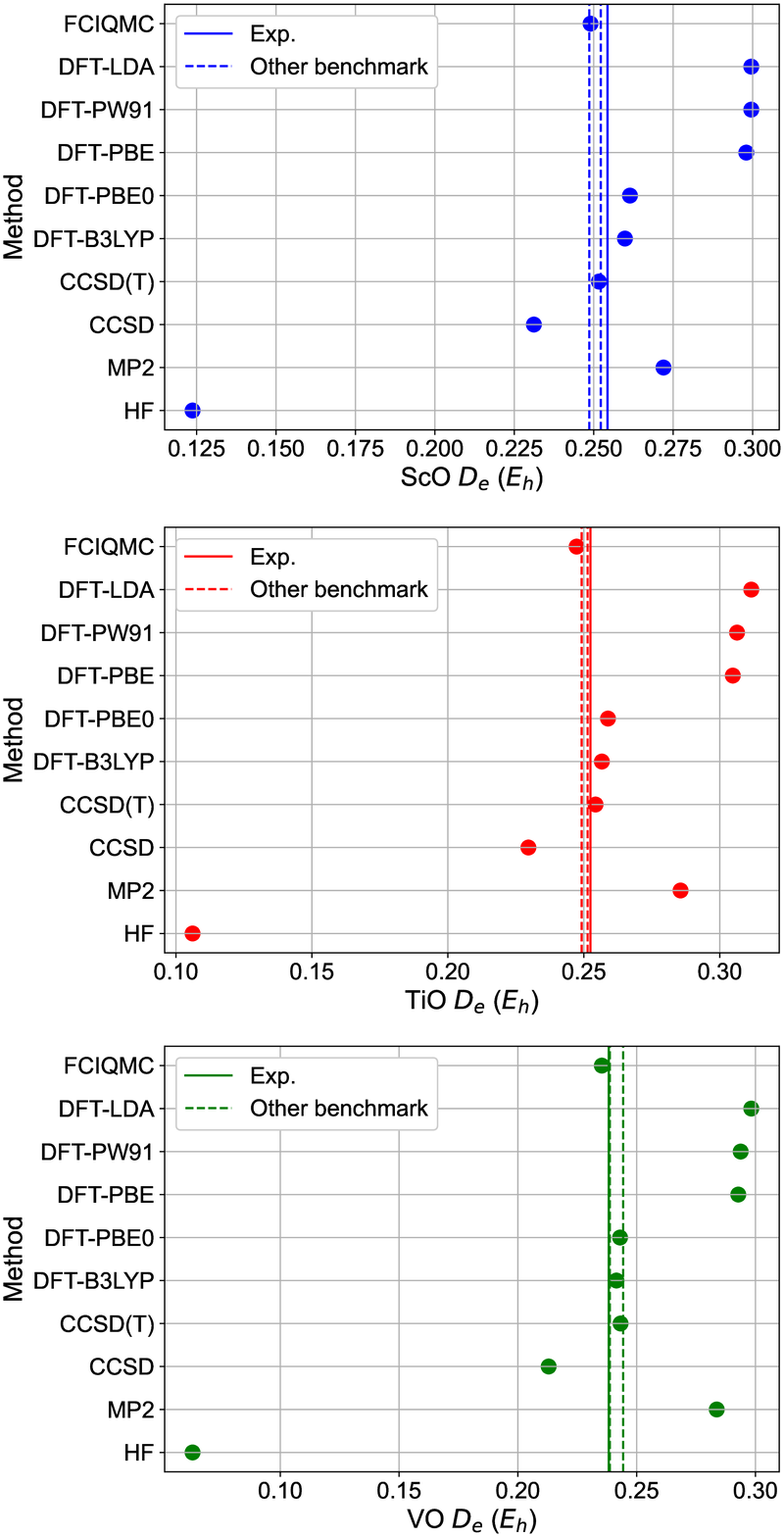}
    \caption{Dissociation energy of ScO $^2\Sigma^+$, TiO $^3\Delta$ and VO $^4\Sigma^-$ from different \textit{ab initio} methods, including WFT (HF, MP2, CCSD, CCSD(T)) and DFT (B3LYP, PBE0, PBE, PW91, LDA) methods. The basis set is cc-pVTZ. Core electrons are not frozen. ``Exp" are experiment results from the same source as in Table \ref{Morse_param}.``Other benchmark" are the upper bound and the lower bound of benchmark results with aug-cc-pVTZ basis set, including auxiliary field quantum Monte Carlo (AFQMC); semistochastic heat-bath configuration interaction (SHCI); and self-energy embedding theory (SEET) \cite{williams_direct_2020}. }
    \label{Fig_De}
\end{figure}

With FCIQMC, we can also compute the ground state 
bonding curve of ScO, TiO and VO.
The dissociation energy is directly calculated as $D_\text{e} = E(\text{M})+E(\text{O})-E(\text{MO})$. 
The Morse parameters can be associated with other bonding parameters, such as the equilibrium bond length ($r_\text{e}$), the total energy at equilibrium geometry ($E_\text{eq}$), and the harmonic frequency ($\omega$), as is shown in Table \ref{Morse_param}.

In ScO $^2\Sigma^+$, TiO $^3\Delta$ and VO $^4\Sigma^-$, the dissociation energies agree quite well with experiment, with an underestimation of approximately $5\text{mE}_\text{h}$, $5\text{mE}_\text{h}$, and $2\text{mE}_\text{h}$, respectively.
We note that in all cases the relative error of FCIQMC from experiment is around or less than 2\%. 
%
%
The equilibrium bond length and harmonic frequency also reproduce experimental data quite well.
For example, in ScO, the deviations of FCIQMC from experiment are 0.4pm and $51\text{cm}^{-1}$, respectively. 
Our AS-FCIQMC results are also in line with other benchmark calculations reported in Ref. \onlinecite{williams_direct_2020}.

The dissociation energy from other WFT and DFT methods are shown in Fig. \ref{Fig_De}.
In all WFT and DFT methods tested, cc-pVTZ basis set is used. 
Different from the relatively small error in excitation energy (e.g. error $<5\text{mE}_\text{h}$ in ScO $^2\Delta$ and $^2\Pi$, Fig. \ref{Fig_ScO}(e)), HF, MP2 and CCSD predict quite inaccurate $D_e$.
CCSD result is lower by about $18\text{mE}_\text{h}$, MP2 result is higher by about $23\text{mE}_\text{h}$, and HF result is much too low.
CCSD(T) is the most accurate one among the WFT methods tested, with difference from FCIQMC less than $3\text{mE}_\text{h}$.
DFT calculations with the two hybrid functionals, B3LYP and PBE0, perform much better for the dissociation energy than the excitation energy. 
They are consistent with FCIQMC result with overestimation around $10\text{mE}_\text{h}$.
PBE, PW91 and LDA are less accurate, with deviation from FCIQMC over $50\text{mE}_\text{h}$. 
TiO $^3\Delta$ and VO $^4\Sigma^-$ cases are both similar to ScO $^2\Sigma^+$. 
It is worth noting that in VO $^4\Sigma^-$ B3LYP and PBE0 functionals in DFT are as accurate in $D_e$ as CCSD(T).
Their deviations from FCIQMC all fall within $6\sim8\text{mE}_\text{h}$.

\section{Discussions}

\begin{figure}
    \centering
    \includegraphics[width = 8.0cm]{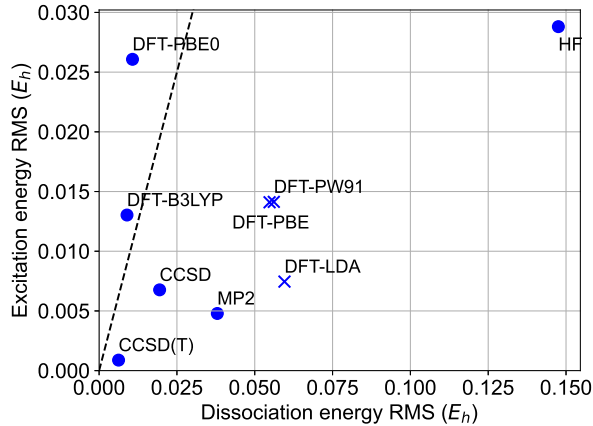}
    \caption{Root mean square (RMS) error of WFT and DFT methods in dissociation energy and excitation energy with respect to AS-FCIQMC. The crosses indicate that there are convergence failures in excitation energy calculations with these methods. The dashed line passes the origin with a slope of 1. 
    }
    \label{disso_excit}
\end{figure}

FCIQMC provides a theoretically accurate evaluation of FCI solution under a certain basis set, and thus can be used as a benchmark for other \textit{ab initio} methods. 
The comparison between AS-FCIQMC and several WFT and DFT methods has been shown in the previous section, and combined comparison is shown in Fig. \ref{disso_excit}.
Overall, our calculations show that WFT theory methods are generally better than DFT in excitation energies, whereas DFT with hybrid functionals may be a good low-cost choice for predicting dissociation energy of transition metal oxide molecules. 
Among WFT methods, CCSD(T) is the most accurate, with excitation energy RMS $<1\text{mE}_\text{h}$, and dissociation energy RMS about $6\text{mE}_\text{h}$.
The error of CCSD and MP2 is less than $7\text{mE}_\text{h}$ in excitation energy, and less than $40\text{mE}_\text{h}$ in dissociation energy. 
Among DFT methods, B3LYP and PBE0 is accurate in dissociation energy within $10\text{mE}_\text{h}$. 
We note that although ScO $^2\Delta$ and TiO $^1\Sigma^+$ are quite multi-configurational according to their most populated configurations' amplitude, their excitation energies are well reproduced by CCSD(T). 
This indicates the possible application of single-configurational CCSD(T) method to multi-configurational systems with 3d elements, such as oxygen vacancies in $\text{TiO}_2$ crystals\cite{Chen_TiO2v_2020}.
Overall, our results suggest that a combination of DFT with hybrid functionals to obtain the ground state structure and CCSD(T) to calculate the excitation energies is a reasonable low-cost choice to examine more complex transition metal oxide molecules and materials.

It is worth noting that, although FCIQMC used in this study is very accurate within the basis set used, the comparison between \textit{ab initio} calculations and experiments may be affected by many other reasons.
At the electronic structure level, errors may come from basis set incompleteness error (BSIE), frozen core approximation, relativistic effect, etc.
As we improve the accuracy of our electronic structure methods, the differences considered with respect to experiments become smaller,
thus the other error causes become more important.
In this study we are satisfied with the size of the computational errors, which are minor 
enough for us to discuss the performance of other commonly used electronic structure methods.
Therefore, we leave the examination of these errors for the future.
Note that for BSIE there are approaches recently combined with FCIQMC to correct BSIE, including 
the explicitly correlated F12 method \cite{Kersten_ExC&G1_2016} and the transcorrelated method \cite{Luo2018}. 

Moreover, our FCIQMC calculations may provide additional insights to the bonding of Sc, Ti, and V oxide materials.
Scandium oxide has a most stable chemical formula of $\text{Sc}_2\text{O}_3$, in which the bonding orbitals are similar to the closed shell orbitals from the atomic orbitals of Sc[Ar] and O[Ne]. 
ScO molecule represents a reduced form of $\text{Sc}_2\text{O}_3$. 
The large excitation energies are in line with the fact that $\text{Sc}_2\text{O}_3$ has a large band gap.
Having the excitation energies underestimated, DFT methods are likely to underestimate the band gap of $\text{Sc}_2\text{O}_3$.
TiO molecule represents a reduced form of $\text{Ti}\text{O}_2$, the most stable form of titanium oxide.
In $\text{Ti}\text{O}_2$, oxygen vacancy is known to be a ubiquitous and important defect, which is similar to TiO molecule that two excess electrons are introduced beyond the closed shell bonding orbitals from Ti[Ar] and O[Ne].
Our FCIQMC calculations indeed confirm relatively small excitation energies and competing excited states.
These features are consistent with the nature of competing defect states in the semiconducting $\text{Ti}\text{O}_2$.
For TiO, DFT does not show a consistent underestimation or overestimation of excitation energy, hence it is difficult to assess which DFT functional could describe well the defect states.
VO is the most complex among the three molecules, having high spin ground states and multiple competing excited states.
Vanadium oxide materials are also more complex with multiple stable valence states ($\text{V}_2\text{O}_3$, $\text{V}\text{O}_2$, $\text{V}_2\text{O}_5$).
Due to such complexities the performance of various methods on vanadium oxides is yet to establish.

\section{Conclusion}

In conclusion, we have carried out systematic calculations on a number of low-lying electronic states of ScO, TiO and VO molecules using AS-FCIQMC and other electronic structure methods.
In particular, we have reported first AS-FCIQMC simulations of the excited states of these molecules, and revealed the configurational population of these states.
Our FCIQMC results set benchmarks for other electronic structure methods and provide insights to their futher developments.
Our study shows that FCIQMC, with continuous developments, is a promising approach to improve our understanding of transition metal oxide materials in the future.

\begin{acknowledgments}
This work was supported by the National Key R\&D Program of China under Grant No.2016YFA030091, the National Natural Science Foundation of China under Grant No. 11974024, and the Strategic Priority Research Program of Chinese
Academy of Sciences under Grant No. XDB33000000.
We are grateful for computational resources provided by Peking University, the TianHe-1A supercomputer, Shanghai Supercomputer Center, and Songshan Lake Materials Lab.
\end{acknowledgments}


%
    

\end{document}